\documentclass[letterpaper,twocolumn,10pt]{article}

\usepackage[latin1]{inputenc}
\usepackage[english]{babel}
\usepackage{scs}
\usepackage{graphicx}
\usepackage{amssymb,amsmath}
\begin{document}

%
%
\title{Investigating Immune System Aging: System Dynamics and Agent-Based Modeling}
\author{
Grazziela P. Figueredo, Uwe Aickelin \\
IMA Research Group, Computer Science School, Nottingham University, Wollaton Road, Nottingham, NG8 1BB UK \\
gzf@cs.nott.ac.uk, uxa@cs.nott.ac.uk}

\maketitle


\begin{abstract}
\noindent System dynamics and agent based simulation models can
both be used to model and understand interactions of entities
within a population. Our modeling work presented here is concerned
with understanding the suitability of the different types of
simulation for the immune system aging problems and comparing
their results. We are trying to answer questions such as: How fit
is the immune system given a certain age? Would an immune boost be
of therapeutic value, e.g. to improve the effectiveness of a
simultaneous vaccination? Understanding the processes of immune
system aging and degradation may also help in development of
therapies that reverse some of the damages caused thus improving
life expectancy. Therefore as a first step our research focuses on
T cells; major contributors to immune system functionality. One of
the main factors influencing immune system aging is the output
rate of naive T cells. Of further interest is the number and
phenotypical variety of these cells in an individual, which will
be the case study focused on in this paper.

\end{abstract}

%
%
\section{Introduction}
\label{sec:introduction}

System dynamics (SD) is an approach to help understand the
behaviour of complex systems over time. It works with feedback
loops, stocks and flows that help describe a system's
nonlinearity. The methodology of developing SD models suggest some
stages that should be accomplished in order to reach the final
model. The first stage is identification of the problem followed
by development of a dynamic hypothesis explaining its causes. The
next stages are building the computational model, testing,
validation and implementation.

Agent based modelling (ABS) is concerned with models in which
agents interact with each other, with a view to observing their
behaviors given changes to the environment.

This work compares SD and ABS immune system aging models that
involve interactions which influence the naive T cell populations
over time. The models are based on the mathematical equations
defined in~\cite{Murray:2003}. In their work, Murray et al.
\cite{Murray:2003} propose a model with a set of equations to fit
observed data and estimate the likely contribution of each of the
naive T cell repertoire maintenance method.

In our work, we converted the original model into simulations and
observe other important points not explored in~\cite{Murray:2003}.
Our objective is to understand the suitability of simulation for
immune system aging problems. Moreover, we want to establish a
comparison between the SD and ABS results, in terms of accuracy
and complexity. In order to understand the problem and how the
model was created, some concepts involving the aging of the immune
system will be presented.

This paper is organized as follows. Section 2 presents the related
work review, section 3 presents the immunological concepts
necessary for the understanding of the modelled problem. Next, in
Section 4, the conceptual and mathematical model are presented.
Sections 5 and 6 present, respectively, the system dynamics and
the agent based models, followed by the results obtained,
conclusions and next steps of this ongoing work.

\section{Related Work}

In this section, we briefly present some of the research
literature related to the comparison of ABS with SD.

In \cite{Wayne:2004}, the authors show the application of both SD
and ABS to simulate non-equilibrium ligand-receptor dynamics over
a broad range of concentrations. They concluded that both
approaches are powerful tools and are also complementary. In
addition, in their case study, they did not indicate a preferred
paradigm, although SD is an obvious choice when studying systems
at a high level of aggregation and abstraction, and ABS is well
suited to studying phenomena at the level of individual receptors
and molecules.

In \cite{Schieritz:2003} the authors present a cross-study of SD
and ABS. They define their features and characteristics and
contrast the two methods. In addition they present ideas of how to
integrate both approaches. The theoretical work presented by
\cite{Pourdehnad:2002} compares ABS and SD conceptually, and
discusses the potential synergy between them in order to solve
problems of teaching decision-making processes.

The work presented in \cite{Stemate:2007} also compares these
modelling approaches and identify a list of likely opportunities
for cross-fertilization. The list presented is not exhaustive and
should be a starting point to other researchers to take such
synergistic views even further.

In our work, we intend to proceed with this investigation using as
case studies simulations related to the aging of the immune
system. In the next section we present concepts necessary to
understand our case study.

\section{Background}
\label{background}

Aging is a complex process that negatively impacts on the
development of the immune system and its ability to
function~\cite{Bulati:2008}. The changes that characterise the
aging of the immune system are called collectively {\it
immunosenescence} or {\it decrease in immunocompetence}.

The decrease of immunocompetence in the elderly can be envisaged
as the result of the continuous challenge of the unavoidable
exposure to a variety of viruses, bacteria, food and other harmful
substances ~\cite{Franceschi:2000}. These exposures cause
persistent life-long immune system stress, responsible for filling
of the ``immunological space'' by an accumulation of immune cells
and immune memory cells~\cite{Franceschi:2000}.

With age, there is also a significant reduction of certain
specific kinds of immune cells, such as naive T cells, caused by
the shrinkage of the organ that generates them, i.e. the thymus.
Naive T cells are able to respond to novel pathogens that the
immune system has not yet encountered. Lower numbers of these
cells eventually leave the body more susceptible to infectious and
non-infections diseases~\cite{Murray:2003}.

Before 20 years of age, naive T cells are sustained primarily from
thymic output~\cite{Murray:2003}. However, in middle age there is
a change in the source of naive T cells; as the thymus involutes,
there is considerable  reduction in its T cell output. Thus, new T
cells are mostly produced by peripheral expansion. There is also a
belief that some memory cells have their phenotype reverted back
to naive cells~\cite{Murray:2003}. Over time, the repertoire of
naive T cells shrinks proportionately to faced threats, while
memory increases~\cite{Murray:2003} \cite{Franceschi:2005}
\cite{Franceschi:2000}.

Thus, late in life, the T cell population becomes less diverse and
a small number of antigen-specific T cell clones can grow to a
great percentage of the total T cell population. This takes over
the space needed for other T cells, resulting in a less diverse
and ineffective immune system. Eventually, there might be not
enough naive T cells left to mount an effective defense and the
immunological space is filled with memory cells.

The aim of this work is to develop models considering each of the
above characteristics in order to investigate if these simulation
techniques are suitable for investigating the immunosenescence
phenomenon. De Martinis~\cite{Franceschi:2005} and
Franceschi~\cite{Franceschi:2000} state that the most important
characteristics of immunosenescence are the accumulation of memory
T cells, reduction of naive T cells, shrinkage of the thymus and a
filling of immunological space.

As we can see from the above, one of the main factors in the
process of immunosenescence is the number and phenotypical variety
of naive T cells in an individual, which changes with age in
quantity and diversity. Recent
research~\cite{Cancro:2009}\cite{Dowling:2009}\cite{Kovacs:2009}
suggests that population-based models of T-cell repertoire
evolution may guide new developments for treating diseases and
help the recovery of the system after depletion caused by
infections, radiation and age. These factors influenced the choice
of the simulation case study developed in this research. Some work
\cite{Thorn:2007} has been done in simulation to understand
certain aspects of immunology and other biological tissue
patterns. However relatively little has been done in terms of
immune degradation. Hence, our conceptual model considers
shrinkage of the thymus and depletion of naive T cells, which will
be explained in the next section.

\section{The Model}
\label{model}

\subsection{Naive T Cell Output}

Thymic contribution in an individual are quantified by the level
of a biological marker called `T cell receptors excision circle'
(TREC). TREC is circular DNA originated during the formation of
the T-cell receptor. The percentage of T cells possessing TRECs
decays with shrinkage of thymic output and activation and
reproduction of naive T cells~\cite{Murray:2003}. This means that
naive T cells originating from the thymus have a greater
percentage of TREC than those originating through other
proliferation.

Our model proposed here is based on data and equations obtained
in~\cite{Murray:2003}, which is concerned with establishing an
understanding naive T cell repertoire dynamics. The objective of
the model is to determine the likely contribution of each of the
naive T cell's sources by comparing estimates of the presence of
TREC in the cells (see Figure 1). The dynamics of the sustaining
sources, i.e. naive proliferation, TREC and reversal of memory to
naive T cells are modelled mathematically.

\begin{figure}[!htpb]
 \begin{center}
  \resizebox{8cm}{!}{\includegraphics{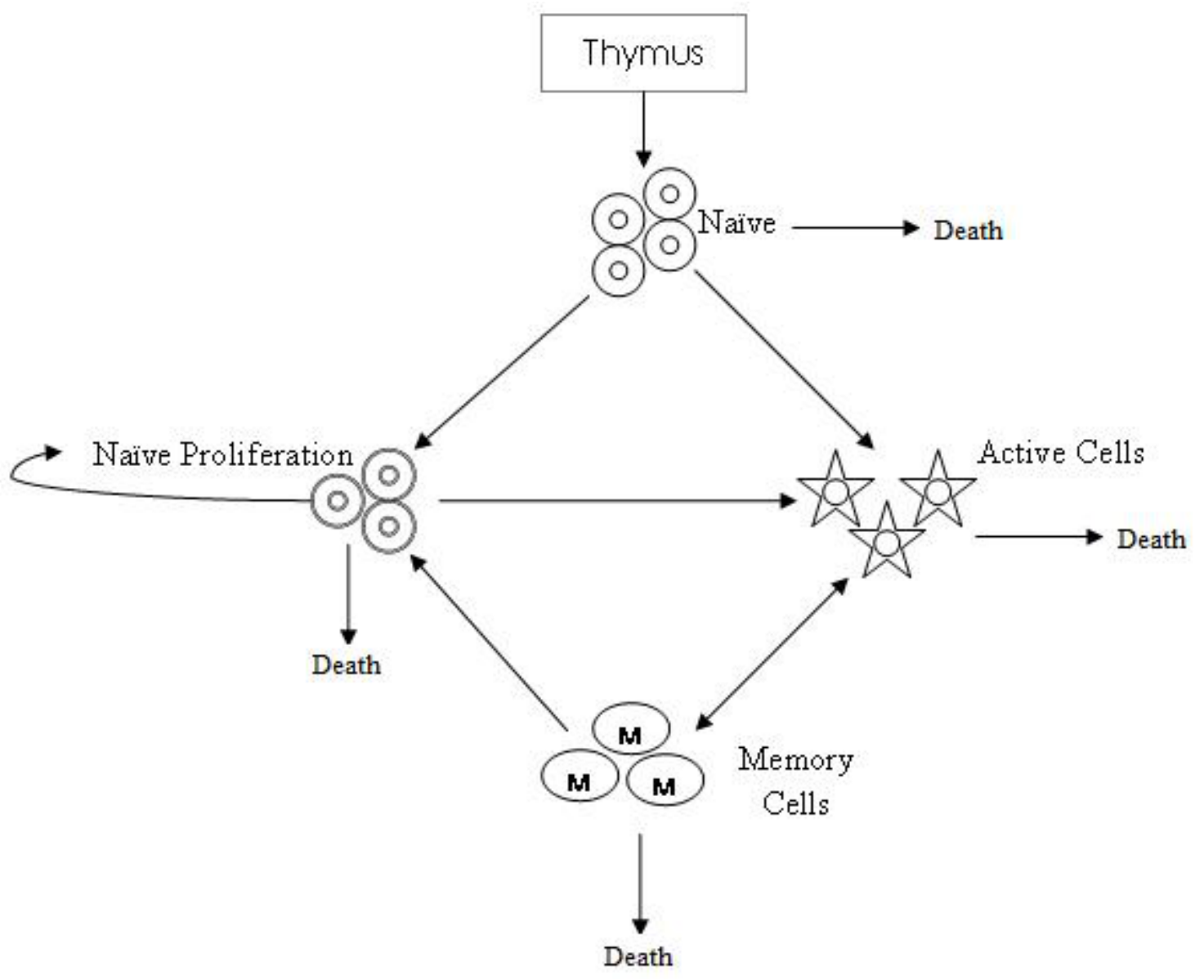}}
 \end{center}
 \label{fig:model}
 \caption{Dynamics of Naive T cells.}
\end{figure}

\subsection{The Mathematical Model}
\label{mathematical}

The mathematical model proposed in~\cite{Murray:2003} is described
by equations (1) to (6) below. In these equations, $N$ is the
total number of naive cells of direct thymic origin, $N_{p}$ is
the number of naive cells that have undergone proliferation, $A$
is the number of activated cells, $M$ is the number of memory
cells and $t$ is time (in years). The first differential
equation is:

$$ \frac{dN}{dt} = s_0e^{-\lambda_tt}s(N_p)-[\lambda_n+\mu_ng(N_p)]N, \ \ \ \ \ \ \ (1)$$
where: $s_0$ is the thymic output; $\lambda_t$ is the thymic decay
rate; $s_0e^{-\lambda_tt}s(N_p)$ represents the number of cells
that arise from the thymus where $s(N_p)$ is the rate of export of
the thymus defined by:

$$ s(N_p) = \frac{1}{1+\frac{\bar{s}^{N_p}}{\bar{N_p}}}\ \ \ \ \ \ \ (2)$$

Also, $\lambda_nN$ represents the naive cells' incorporation into
the naive proliferating pool where $\lambda_n$ is the naive
proliferation rate; $\mu_n$ is the thymic naive cells death rate;
$\mu_ng(N_p)N$ represents the naive cell death rate where the
function $g(N_p)$ is the death rate of between naive TREC-positive
and naive TREC-negative, defined as:

$$ g(N_p) = 1+ \frac{\frac{bN_p}{\bar{N_p}}}{1+\frac{N_p}{\bar{N_p}}}, \ \ \ \ \ \ \ (3)$$

$\bar{N_p}$ and $\bar{s}$ are equilibrium and scaling values
respectively. The second differential equation is:

$$ \frac{dN_p}{dt} = \lambda_nN + [ch(N,N_p) - \mu_n]N_p + \lambda_{mn}M \ \ \ \ \ \ \ (4)$$
where: $c$ is the proliferation rate; $ch(N,N_p)N_p$ represents
the naive proliferation where $h(N,N_p)$ is the dilution of
thymic-naive through proliferation defined by:

$$ h(N,N_p) = \frac{1}{1+\frac{N+N_p}{\bar{N_p}}} \ \ \ \ \ \ \ (5)$$

$\mu_nN_p$ is the death rate of proliferation-originated naive
cells and $\lambda_{mn}$ is the reversion rate from memory into
$N_p$. The final differential equation is:

$$ \frac{dM}{dt} = \lambda_aA - \mu_mM - \lambda_{mn}M. \ \ \ \ \ \ \ (6)$$
where: $\lambda_a$ is the reversion rate into memory and $\mu_m$
is the death rate of memory cells. The parameter values for the
model's can be seen in Table 1.

\begin{table}[ht!]
 \centering
\begin{tabular}{|l|l|}
\hline
rate                 &  value(s)  \\
\hline \hline
$\lambda_t$ & $\frac{log(2)}{15.7}(year^{-1})$ \\
\hline
$\lambda_n$ & 0.22, 2.1, 0.003 \\
\hline
$\mu_n$ & 4.4 \\
\hline
$c$ & 0 (no proliferation) or $\mu_n(1+ \frac{300}{\bar{N_p}})$ \\
\hline
$\lambda_{mn}$ & 0 \\
\hline
$\mu_m$ & 0.05\\
\hline
$\lambda_{a}$ & 0 \\
\hline
\end{tabular}
 \label{Tab:rates}
 \caption{Rate values for the mathematical model.}
\end{table}

For the mathematical model and subsequent simulations, $s0 =
56615$ and the active cells were obtained by a look-up table based
on data collected by \cite{Comans-Bitter:1997}. This table
contains the number of activated CD4 cells per $mm^3$ for early
years. This table was used as a stock for the active cells. From
the active cell stock we update the values of the memory cell
stock.

Equations (1) to (6) will be incorporated in the SD and ABS models
in order to investigate if it is possible to reproduce and
validate the results obtained in~\cite{Murray:2003}. Moreover,
variations of the ratio variables are investigated to understand
the importance of each individual integrand in the system. For
example, it is important to establish how much the proliferation
rate impacts the depletion of naive T cells over age; also, at
which point in time can the system be defined as losing
functionality.

\section{The System Dynamics Model}
\label{SD}

The system dynamics model objective is to simulate the processes
involved in the maintenance of naive T cells. The model was built
taking into consideration all the interactions described by the
mathematical equations defined in the previous section. The
simulation was built using AnyLogic 6.5 University Edition
\cite{AnyLogic:2010}.

The naive T cells, naive T cells from proliferation and memory
cells are stock variables. The stock variable that represents the
number of naive T cells is subject to thymic output, proliferation
and death rates, which are flow variables. The flows between naive
cells and active cells are not defined in the mathematical model.
Therefore we assume that these flows only interfere on the stock
of active cells, which will not be considered in the SD model. The
number of active cells, which is a stock, will be given by a look
up table  containing real values of active cells in the organism.
The graphical representation of the stock of naive T cells and its
flows, which corresponds to Equation 1 on the mathematical model,
can be seen in Figure 2.

\begin{figure}[!htpb]
 \begin{center}
  \resizebox{3.5cm}{!}{\includegraphics{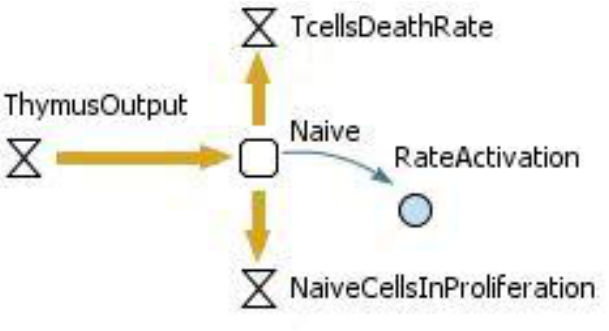}}
 \end{center}
 \label{fig:naive}
 \caption{The naive T cell stock variable and its flows: thymic output, proliferation, activation and death.}
\end{figure}

The amount of naive cells from proliferation is changed by death,
proliferation and reversion from memory rates, according to
Equation 2. The memory repertoire is modified by death and
reversion to a naive phenotype (Equation 6). All the stocks put
together form the SD model and can be seen in Figure 6.

\begin{figure}[!h]
 \begin{center}
  \resizebox{3cm}{!}{\includegraphics{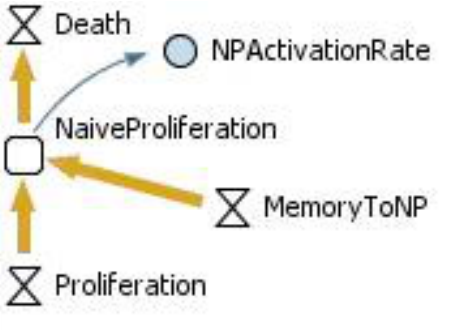}}
 \end{center}
 \label{fig:np}
 \caption{The naive T cells from proliferation stock variable and its flows: death, activation, proliferation and memory reverted to naive.}
\end{figure}

\begin{figure}[!h]
 \begin{center}
  \resizebox{4cm}{!}{\includegraphics{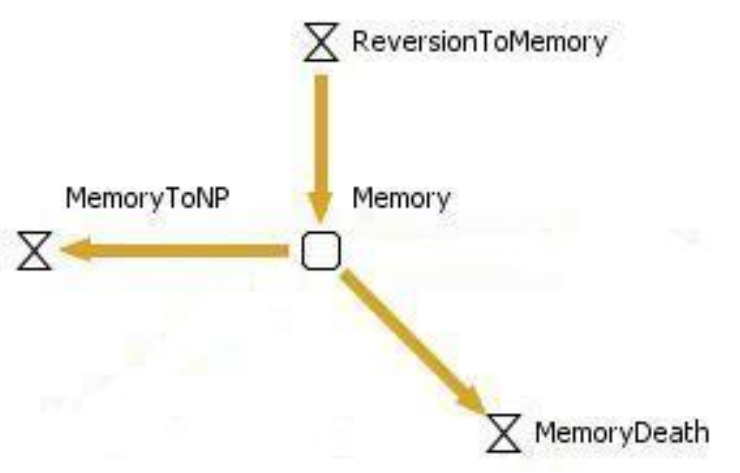}}
 \end{center}
 \label{fig:memory}
 \caption{The memory T cells stock variable and its flows: death and reversion to a naive
phenotype.}
\end{figure}

\begin{figure}[!h]
  \centering
    \resizebox{8cm}{!}{\includegraphics{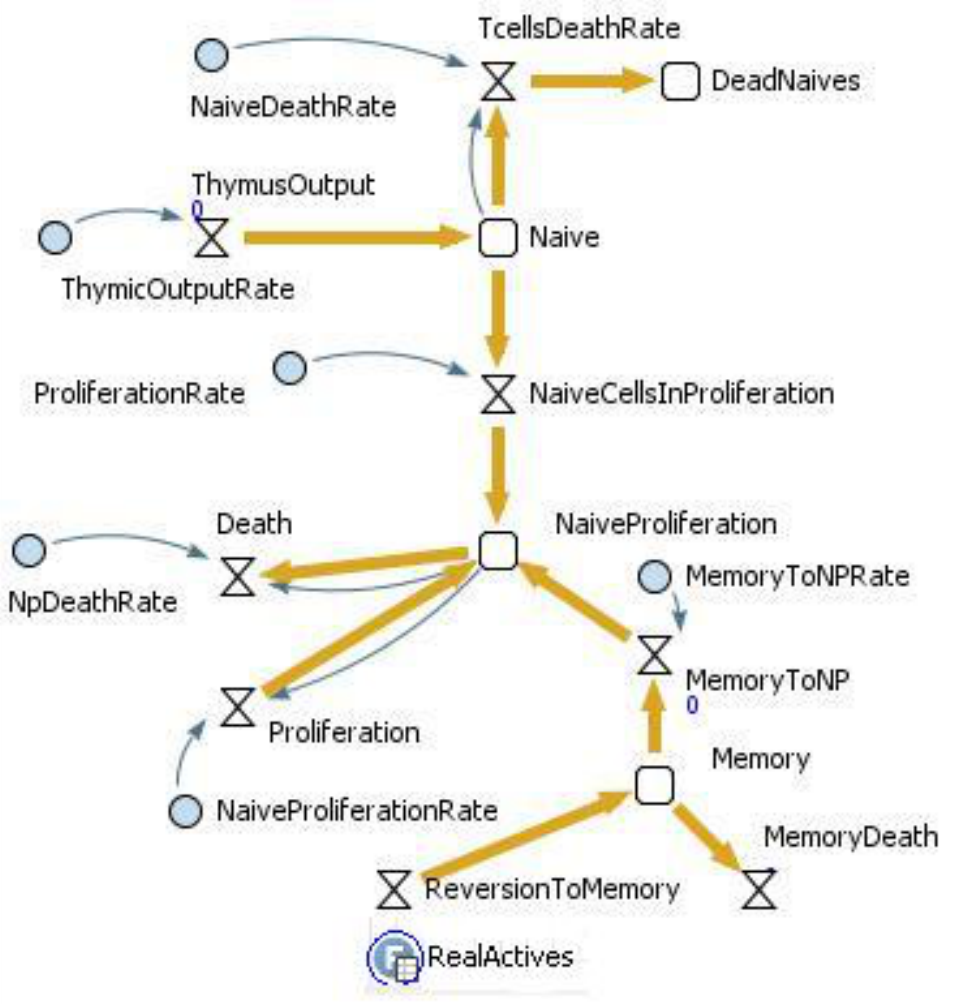}}
  \caption{The system dynamics model's functions and parameters. The function RealActives returns the values for active CD4+ cells at a certain time. The parameters NPActivationRate, NPActivation, NPDeathRate, MemoryToNP and MemoryToNPRate correspond to naive cells from proliferation (NP).}\label{fig:modelSD}
\end{figure}

\section{The Agent Based Model}
\label{ABS}

The SD model was then converted into an ABS model shown in Figure
7. T cells are the agents and the state chart represents all the
stages these agents can assume, i.e. naive, naive from
proliferation, active or memory cells. The agents' state changes
and their death rates are given by the ratios defined in the
mathematical model. Initially all the agents are in the {\it
naive} state. As the simulation proceeds they can assume other
stages according to the transition pathways defined in the state
chart. The agents also reproduce, and the newborn agents which are
also T cells, should assume the same state as their original
agent.

At the beginning of the simulation, there are only agents in the
{\it Naive} state generated by thymic output. From the {\it Naive}
state, the agents can proliferate and go to the {\it
NaiveProliferation} state. If the T cells in the {\it
NaiveProliferation} state reproduce according to the {\it
NaiveProliferationRate}, more agents will be generated in the same
state ({\it NaiveProliferation}) as their progenitors. The same
thing happens to the state {\it Memory}. The agents can also die
according to specific rates, determined by the mathematical model.
The agents in this simulation respond to changes in time and do
not interact with each other directly. Therefore, what we intended
to get from this simulation is the response of each individual
cell to the proposed environment.

\begin{figure}[!h]
 \begin{center}
  \resizebox{8.2cm}{!}{\includegraphics{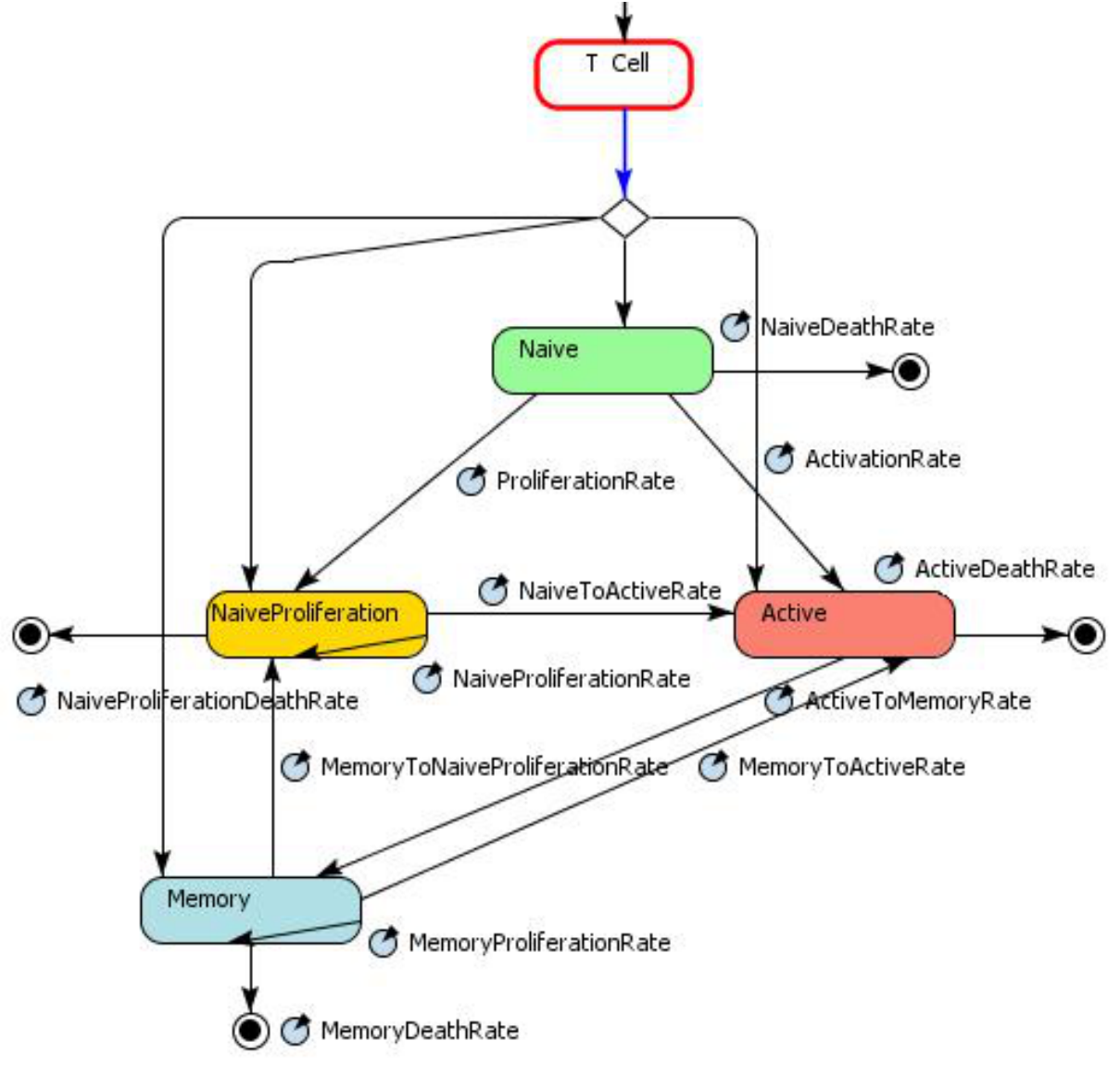}}
 \end{center}
 \label{fig:modelABS}
 \caption{The ABS model. In this simulation, The number of active agents
is also given by a look up table.}
\end{figure}

\begin{center}
\begin{table*}[ht]
{\small \hfill{}}
\begin{tabular}{|l|l|l|}

\hline
Scenario                    & Description       & Parameters  \\
\hline \hline
 1 & No peripheral proliferation & $c=0, \lambda_n=0.22, \lambda_{mn}=0.05, \overline{N}_p=387, \overline{s}=0.48, b=3.4, \mu_Np=0.13 $ \\

\hline

2 & No homeostatic reduction in thymic export, & $\overline{s}=0,
b=0, \lambda_n=2.1, \lambda_{mn}=0, \overline{N}_p=713$
\\&no homeostatic alteration of naive death rate& \\

\hline

3 & Homeostatic alteration of naive death rate & $\overline{s}=0,
\lambda_n=0.003, \lambda_{mn}=0, \overline{N}_p=392, b=4.2$\\ &
but not thymic export & \\

\hline

\end{tabular}
\label{Tab:models} \caption{Simulation parameters for different scenarios.}
\end{table*}
\end{center}

\section{Experiments}

We study three simulation scenarios defined by~\cite{Murray:2003} with
different values for the parameters. A summary of
parameters changed for each scenario can be seen in Table 2.

One of the case study simulation's objectives is to investigate if
there is the need for naive peripheral proliferation throughout
life. Therefore, for the first scenario the naive peripheral
proliferation rate is set to zero. It also considers reversion
from memory to a naive phenotype.

The second scenario assumes peripheral proliferation with a bigger
rate of naive cells becoming naive proliferating cells (2.1):
There is no reversion from memory to a naive phenotype and no
homeostatic reduction in thymic export. The functions $s$, $g$ and
$h$ from the mathematical model are responsible for controlling
the thymic export, naive death rate and naive peripheral
proliferation, respectively. In order to alter the thymic export,
the parameters $\overline{s}$ and $\overline{N}_p$ have been
changed. The parameter $b$ was set to zero so that the function
$g$ would remain constant during all the simulation and so the
death rate of naive cells.

The third scenario alters the function $g$ over time by setting the
parameter $b$ greater than zero ($b=4.2$). This means that the
death rate of naive T cells from thymus will increase along the
years as the number of naive from peripheral proliferation
increases. There is no change of the thymic export, no reversion
from memory to a naive phenotype and the conversion rate of naive from thymus to naive proliferation is low (equal to 0.003).

The data used for validation of the mathematical model and the
simulations is displayed in Table 3. The data set contains
information about the TREC marker in individuals grouped in age
ranges. The first column of Table 3 shows the age range of the
individuals; the second column has the mean
$\frac{\log_{10}TREC}{{10^6}} PBMC$ (peripheral blood mononuclear
cell) and the third column contains the number of individuals in
each age range. The total number of individuals in the experiment
was $506$. The graphic containing the TREC data (naive from
thymus) and total naive cell data provided by
\cite{Cossarizza1996173} is shown in Figure 8.

\begin{table}[h!]
 \centering
\begin{tabular}{|l|l|l|}
\hline
Age                   &  $\frac{\log_{10}TREC}{{10^6 \times n} PBMC}$      & number of individuals  \\
\hline \hline
0   & 5.03 & 48 \\
\hline
1-4 & 4.93 & 53 \\
\hline
5-9 & 4.86 & 19 \\
\hline
10-14 & 4.86 & 19 \\
\hline
15-19 & 4.56 & 33 \\
\hline
20-24 & 3.88 & 26 \\
\hline
25-29 & 3.75 & 47 \\
\hline
30-34 & 3.61 & 65 \\
\hline
35-39 & 3.54 & 73 \\
\hline
40-44 & 3.52 & 52\\
\hline
45-49 & 3.37 & 55\\
\hline
50-54 & 3.17 & 16\\
\hline
\end{tabular}
 \label{Tab:dataset}
 \caption{The dataset used for validation obtained in \cite{Murray:2003}.}
\end{table}

\begin{figure}[h]
 \begin{center}
  \resizebox{8cm}{!}{\includegraphics{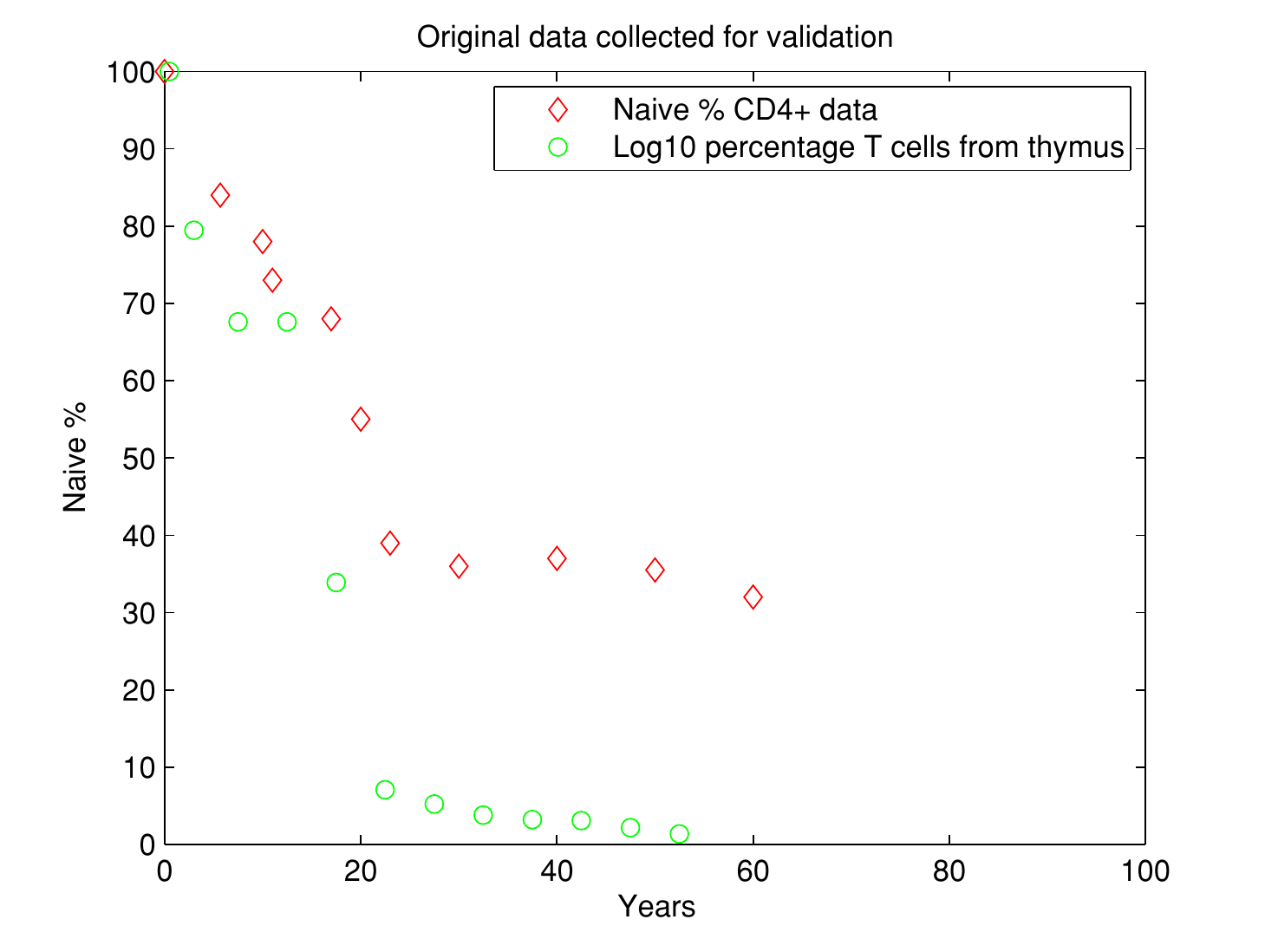}}
 \end{center}
 \label{Results}
 \caption{The dataset used for validation was provided by \cite{Murray:2003} and \cite{Cossarizza1996173}. The data represented by the symbol $\circ$ is the TREC data shown in Table 3 and  $\diamond$ represents the total percentage of CD4+.}
\end{figure}

Each simulation was run for a period of one hundred years
considering the impact of thymic shrinkage per $mm^3$ of
peripheral blood and $3673$ initial naive cells from thymus for
the SD model.

\section{Results} \label{PreliminaryResults}

The SD simulation results for the first scenario can be seen in
Figure 9. The ABS results are shown in Figure 10. The results for
both simulation techniques show a very similar trend curve,
although the ABS results exhibit a more noisy behavior in time.
For this first scenario, the simulation results did not fit the
original data. The naive cells from thymus curve produced shows a
big decay in thymic export on the beginning of life because of the
high death rate.

After the twenties, an exponential decay of thymic export is
observed and the dynamics follows the thymic decay rate rule
defined in the mathematical model. The naive proliferation curve
increases with the decrease of naive from thymus, but as there is
no proliferation of peripheral cells, they die with no
replacement. Thus they follow the same pattern of their only
source, i.e. thymic naive cells. The results indicate that
peripheral proliferation is important for maintenance of naive T
cells.

For scenario $2$, results match the original data more closely
than in the previous scenario. In this case peripheral
proliferation is considered as well as a high rate of naive cells
from the thymus turning into peripheral naive cells. Figure 11
shows the results for SD and Figure 12 for ABS. The naive from
thymus curve shows a big decay in the beginning of life because of
the death and proliferation rates. On the other hand, the naive
from proliferation curve increases with the decrease of the naive
from thymus curve. This pattern is controlled by the $g$ function.

The main difference between these results and results form the
previous scenario is that the amount of naive cells from
proliferation reaches a stable value after the age of twenty with
no further decay. The results indicate the importance of
peripheral expansion, but the need of a smaller rate of naive to
peripheral naive conversion. Moreover, reversion from memory is
not important.

Scenario $3$ takes into account the results produced in the
previous scenarios and adjusts the parameters in a way that a more
accurate output is obtained. The results for SD and ABS can be
seen in Figures 13 and 14, respectively. The naive from thymus
curve presents a decay at the beginning of life followed by an
interval of stability ruled by $s0$. By the age of twenty the
thymic export decreases in an exponential trend. With the decay of
naive from thymus, the naive repertoire changes from the thymic
source to peripheral proliferation source. Therefore, by
performing these simulations it is possible to have an idea of how
the decay of naive cells happens over time. The results now
closely match the original data.

\begin{figure}[!h]
 \begin{center}
  \resizebox{8cm}{!}{\includegraphics{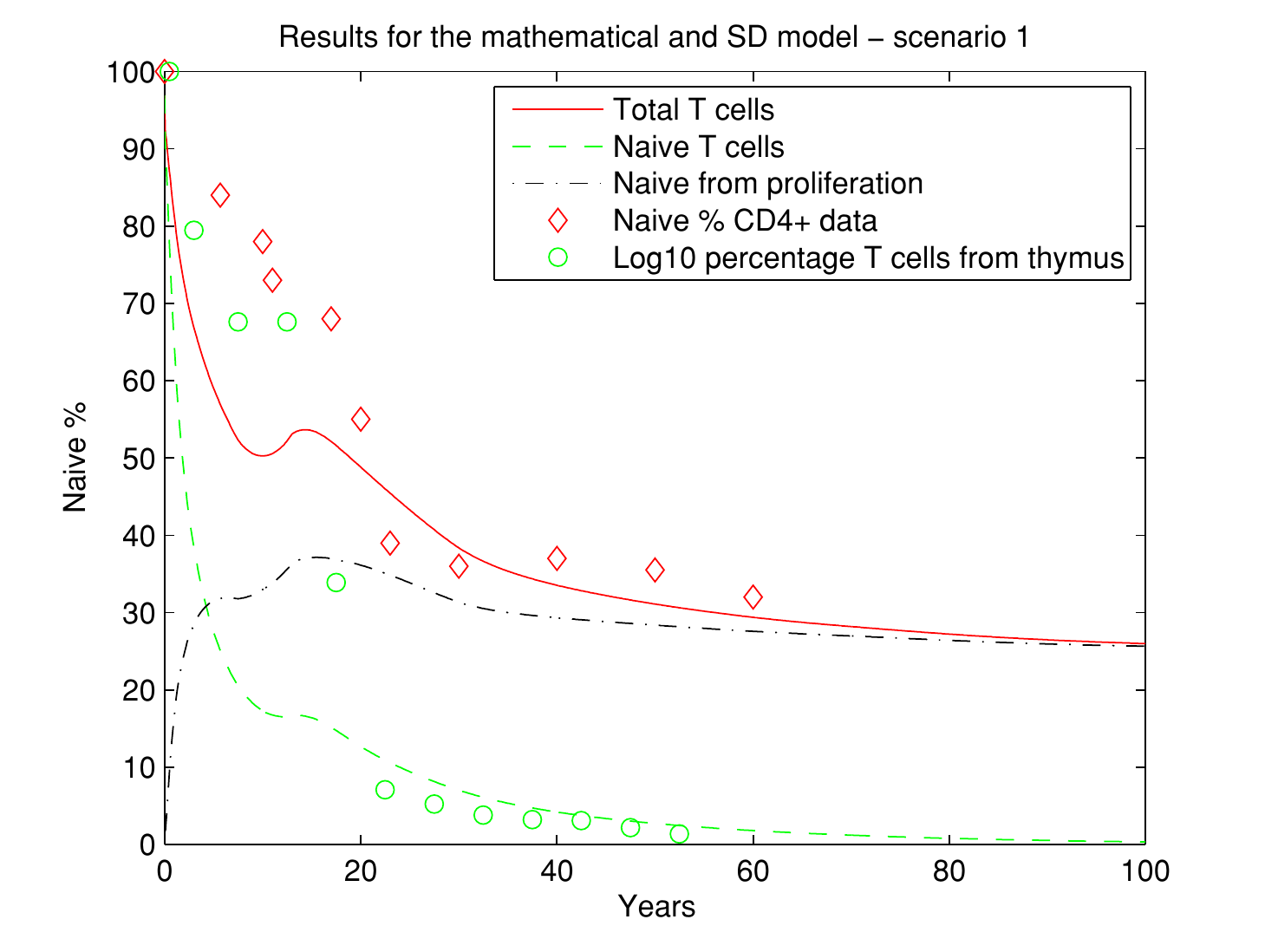}}
 \end{center}
 \label{fig:Scenario1}
 \caption{Results for scenario 1 with SD.}
\end{figure}

\begin{figure}[!h]
 \begin{center}
  \resizebox{8cm}{!}{\includegraphics{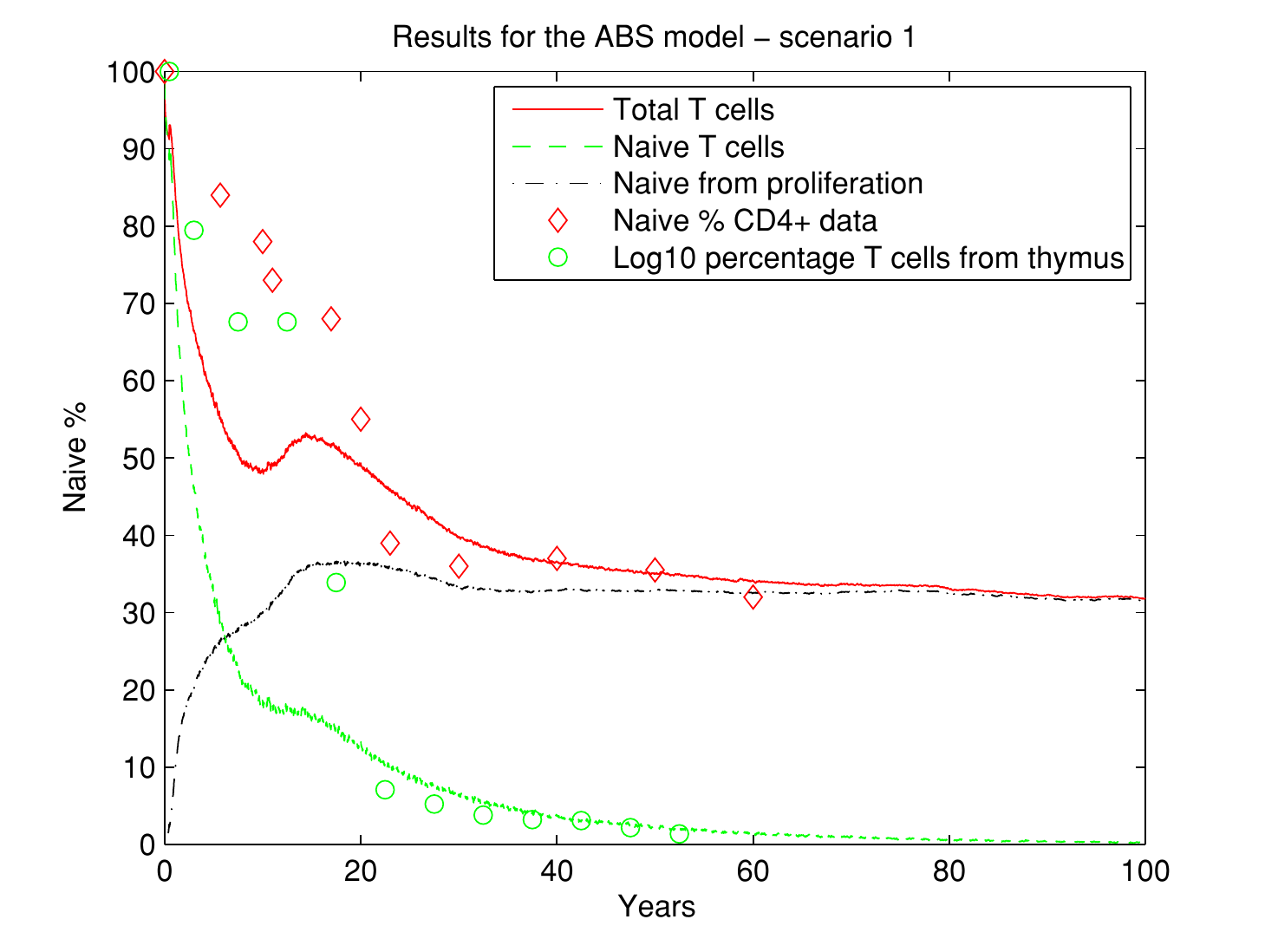}}
 \end{center}
 \label{fig:Scenario1ABS}
 \caption{Results for scenario 1 with ABS.}
\end{figure}

\begin{figure}[!h]
 \begin{center}
  \resizebox{8cm}{!}{\includegraphics{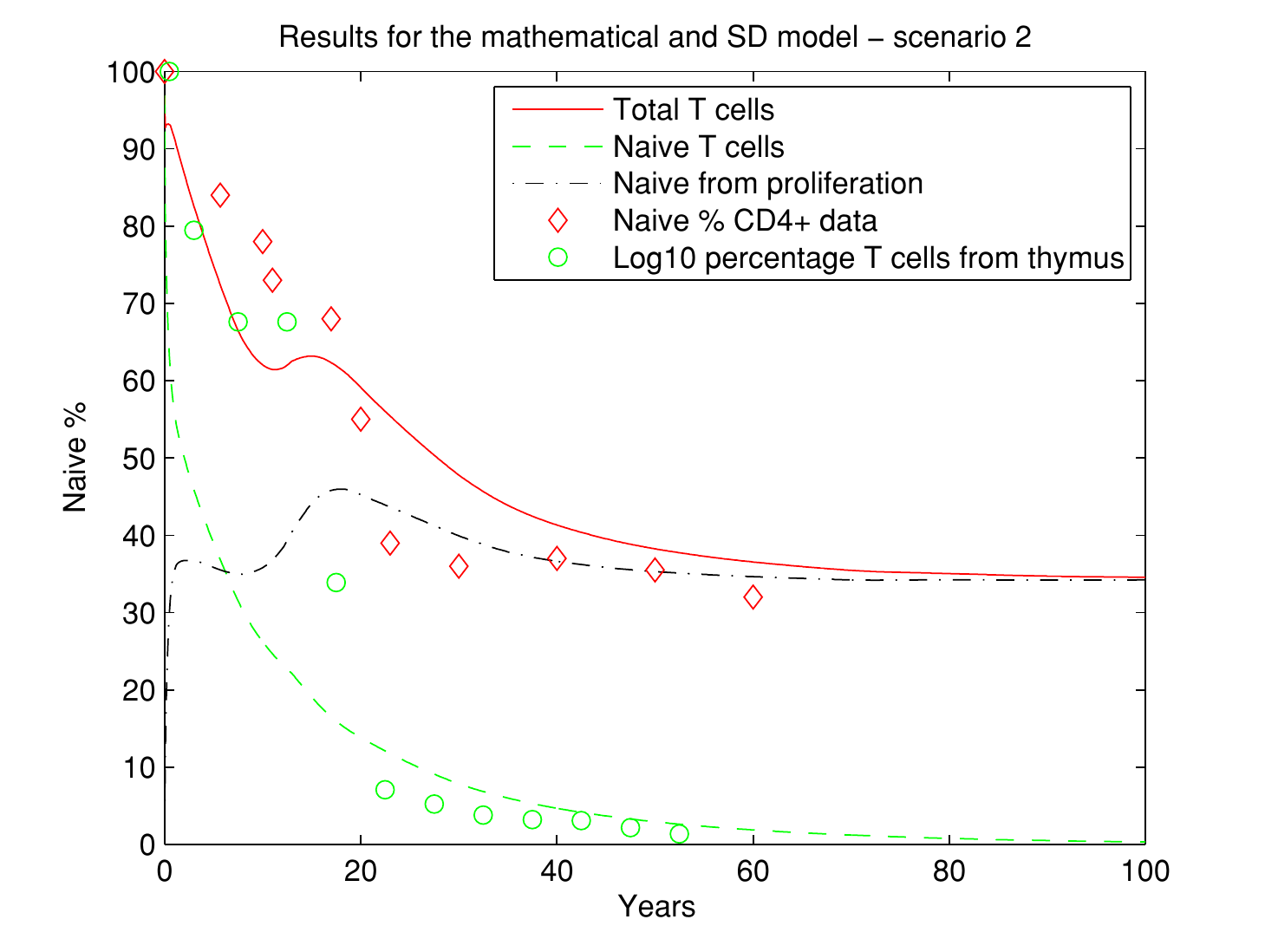}}
 \end{center}
 \label{fig:Scenario2}
 \caption{Results for scenario 2 with SD.}
\end{figure}

\begin{figure}[!h]
 \begin{center}
  \resizebox{8cm}{!}{\includegraphics{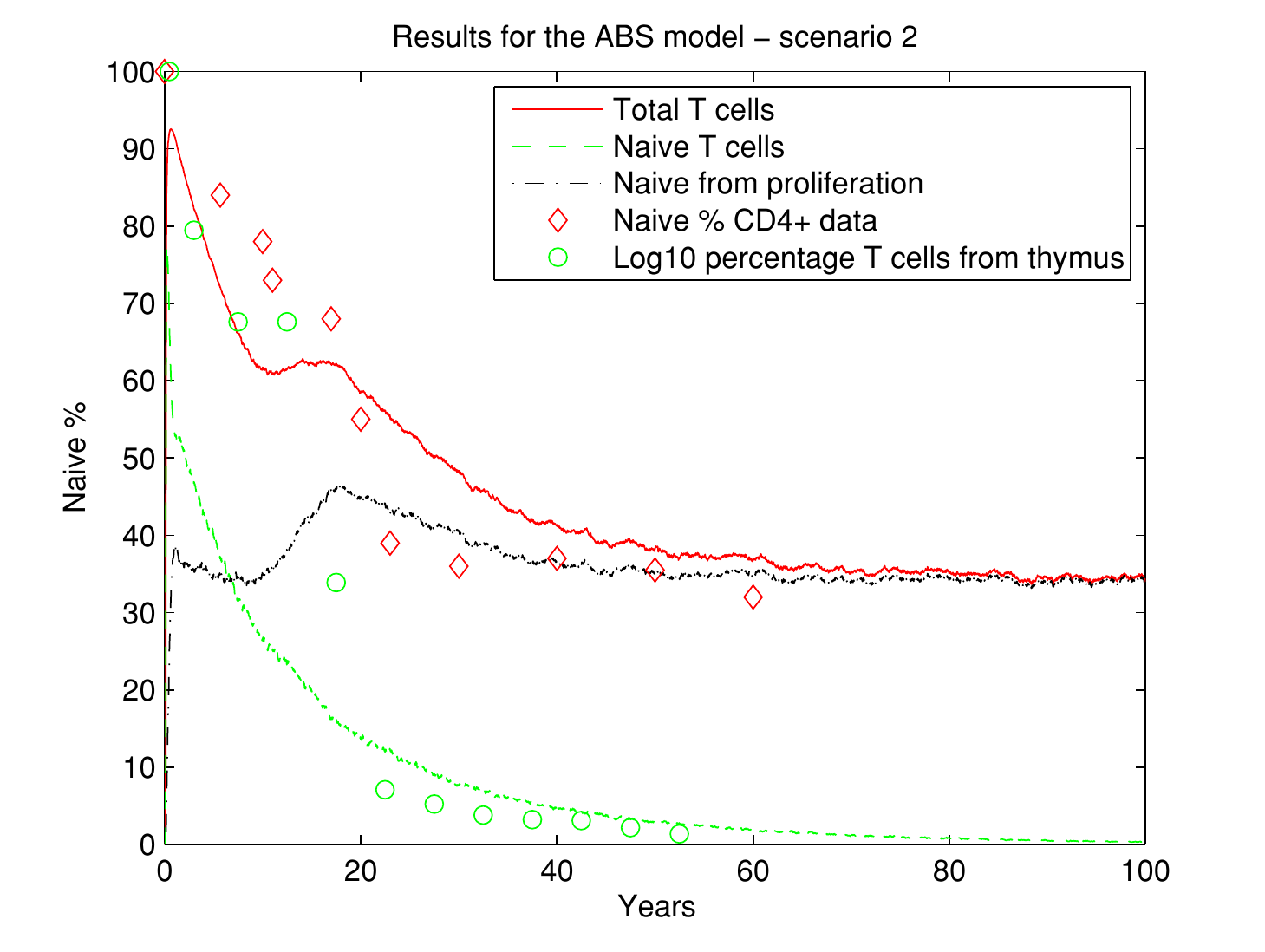}}
 \end{center}
 \label{fig:Scenario2ABS}
 \caption{Results for scenario 2 with ABS.}
\end{figure}

\begin{figure}[!h]
 \begin{center}
  \resizebox{8cm}{!}{\includegraphics{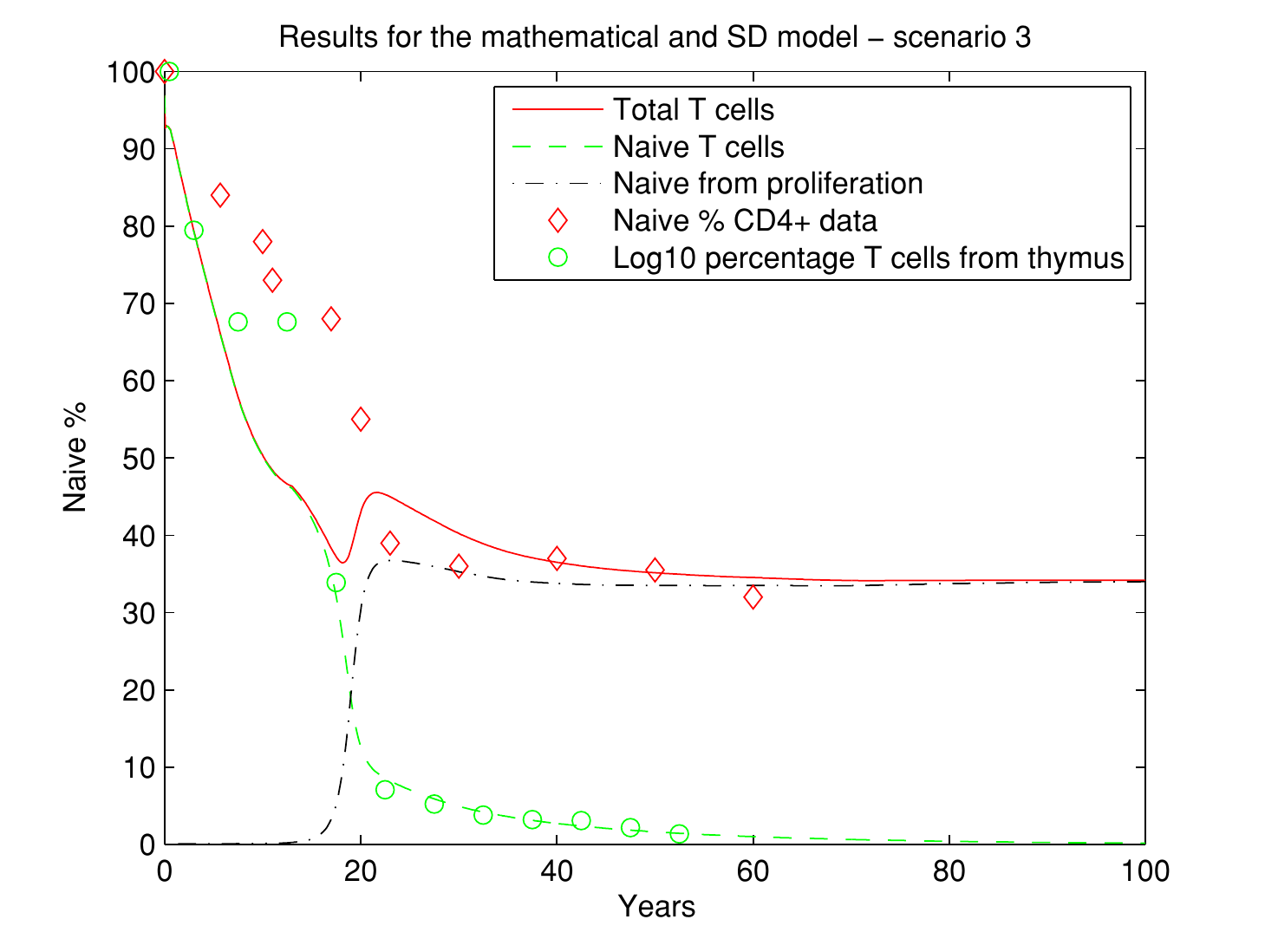}}
 \end{center}
 \label{fig:Scenario3}
 \caption{Results for scenario 3 with SD.}
\end{figure}

\begin{figure}[!h]
 \begin{center}
  \resizebox{8cm}{!}{\includegraphics{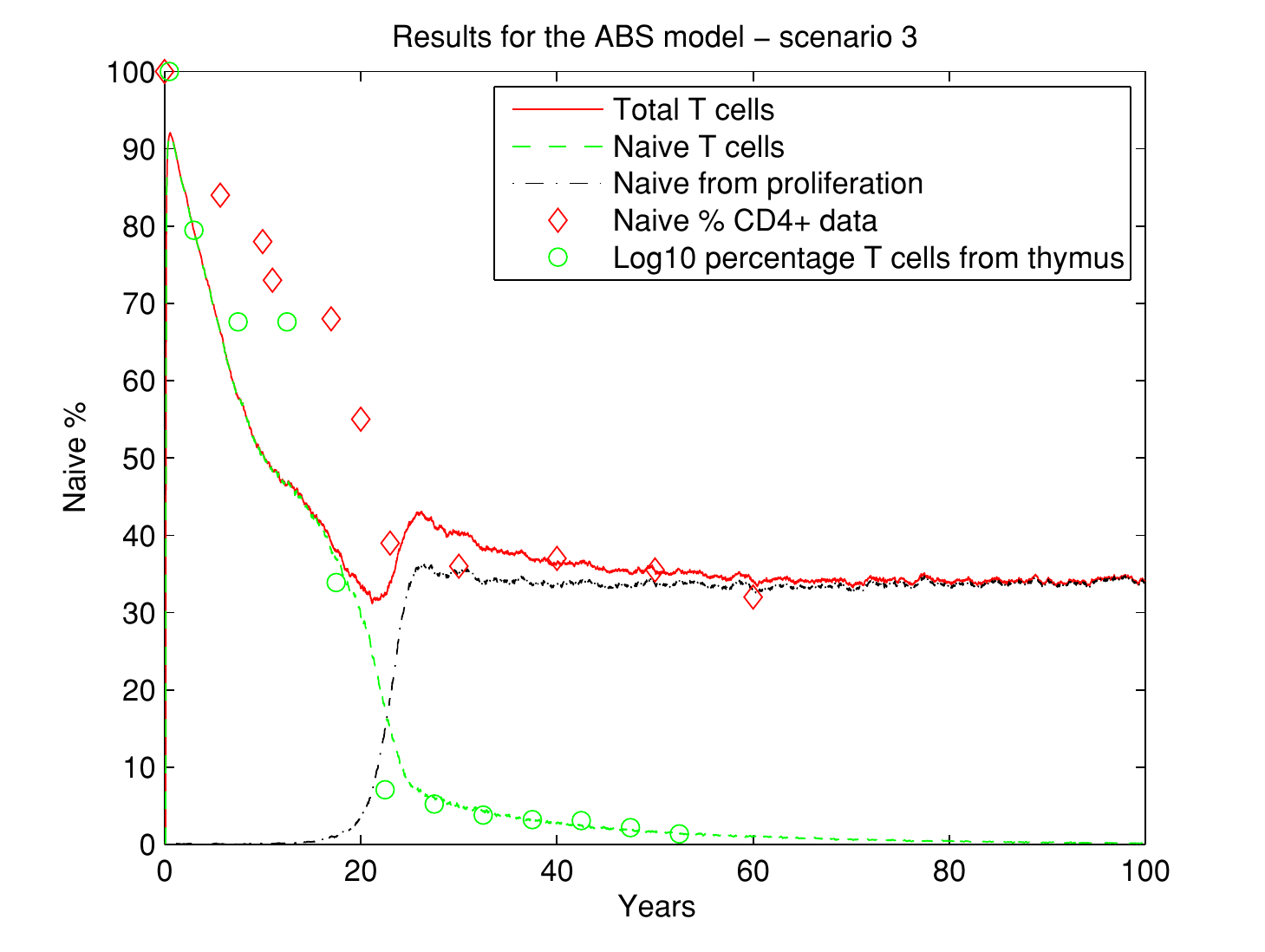}}
 \end{center}
 \label{fig:Scenario3ABS}
 \caption{Results for scenario 3 with ABS.}
\end{figure}

For the three scenarios studied, the simulations produced similar
results for both SD and ABS techniques matching the original data.
Hence, it is possible to conclude that the mathematical model, as
well as immunosenescence problems are suitable for simulation
using both techniques. The differences between the SD and ABS
simulations are in how the conceptual model is represented and
manipulated.

SD gives a systemic view of the conceptual model and tries to
forecast how the system as a whole would evolve in time in an
aggregate manner. This means that each change ratio will be
applied to the entire set of cells. One could argue that SD
represent an idealised version of the system.

In the ABS model, cells are subject to individual rates that occur
during the time slice they were created in. This makes the output
more noisy than the SD results. It also seems to be more
realistic, because in real immune systems, cells have individual
behaviours and responses to the environment. In addition, the ABS
model is easier to adapt to changes in the conceptual model, such
as when change rates occur during a cell's life span or new
cellular interactions between cells.

We also believe that the individual cellular behavior could be
captured in a more simplified mathematical model. For example, the
$s$, $g$ and $h$ functions' control could be incorporated in the
way cells change over time. This would help to further understand
the biological system behavior. On the other hand, SD is simpler
to implement and demands less computational resources such as
memory, processing time and complexity as it can be seen in the
comparison of Tables 4 and 5. The precision measure on the table
is the sum of the square errors (SSE) of the percentages of naive
T cells from thymus. The best results are given by the smaller
numbers of SSE. The other measurements are time in seconds and
memory usage in megabytes. The numbers for the system dynamics are
better in terms of resources consumption as well as precision. We
believe that precision for the ABS was not better because of the
instabilities that occurred in the resulting graphs.

\begin{table}[h!]
 \centering
\begin{tabular}{|l|c|c|c|}
\hline
 Scenario                  & Time (sec) & RAM & SSE \\
\hline

1                          & 0.7        &11& 71696\\

\hline
2                          & 0.8        &11& 5924.4 \\

\hline

3                          & 0.7        &10&  589.9\\
\hline
\end{tabular}
 \label{Tab:compSD}
 \caption{SD processing time, memory usage and precision.}
\end{table}

\begin{table}[h!]
 \centering
\begin{tabular}{|l|c|c|c|}
\hline
 Scenario                  & Time (sec) & RAM (MB)& SSE \\
\hline

1                          &  28800      & 40     & 10349\\

\hline
2                          &   61700          &50      & 20668\\

\hline

3                          &   64800          &   50      &  45834\\
\hline
\end{tabular}
 \label{Tab:compSD}
 \caption{ABS processing time, memory usage and precision.}
\end{table}

\section{Conclusions}

Understanding immunosenescence and its causes may help direct
research into therapies that reverse some of its consequences and
improve life expectancy. The research question raised within this
work focuses on the feasibility of using SD and ABS simulation
tools to understand immune system aging and comparing the two
strategies. The ability to model immune system aging will provide
new means to define a patient-specific `Immune Risk Phenotype'.
Modelling this Risk Type will improve our knowledge of immune
factors that contribute to morbidity and mortality and facilitate
the design of appropriate interventions.

The main factors influencing the process of immunosenescence
include the number and phenotypical variety of naive T cells in an
individual, which change with age in quantity and diversity. At
the beginning of life, the thymus is the principal source of naive
T cells. With age, there is a decay in thymus output and a shift
between the main source of naive T cells. It is believed that the
sustenance of naive T cells in the organism is provided by
peripheral expansion, reversion from a memory phenotype, and
long-lived T cells. Data related to thymic output rate was
collected by~\cite{Murray:2003} and a mathematical model was
build. This mathematical model was used as the baseline for the
development of SD and ABS models.

Three simulation scenarios were studied and the simulation outputs
were broadly similar for both SD and ABS. Hence, it is possible to
conclude that the mathematical model, as well as immunosenescence
problems are suitable for simulation using both techniques. The
differences between the SD and ABS outputs are in how the
conceptual model is represented and manipulated.

The SD based simulation is closer to the underlying mathematics,
but has the disadvantage of being high-level, with complete
homogeneity of simulated entities. On the other hand, ABS allowed
a representation of each entity and heterogeneity. However, ABS
increases the demand for computational resources.

As future work, it is intended to build a framework to develop
immunosenescence simulation models. We also want to define guidelines to help decide
when each simulation type should be used in an immunosenescence
model.

Finally, we are interested to investigate how combining ABS and SD
models could help improving the simulations. As we have shown, immunosenescence can be simulated by more than one simulation type,
however our hypothesis is that a combination of methodologies will
provide higher prediction accuracy given the same amount of input
data and computational resources.

\renewcommand{\baselinestretch}{0.98}
\bibliographystyle{plain}
{\small
\bibliography{scsc_new}}
\renewcommand{\baselinestretch}{1}

\end{document}